\documentclass[prl,aps,twocolumn,superscriptaddress,showpacs,amsmath,amssymb,floatfix]{revtex4}

\usepackage{graphicx}

\newcommand{\be}{\begin{equation}}
\newcommand{\ee}{\end{equation}}

\begin{document}
\title{Quantum fluctuations and collective oscillations of a Bose-Einstein condensate in a 2D 
optical lattice}

\author{G. Orso}
\affiliation{BEC-INFM and Dipartimento di Fisica, Universita' di Trento, 
1-38050 Povo, Italy}
\author{C. Menotti}
\affiliation{BEC-INFM and Dipartimento di Fisica, Universita' di Trento, 
1-38050 Povo, Italy}
\affiliation{ICFO - The Institute for Photonic Sciences,
 Mediterranean Technology Park, \\
 E-08860 Castelldefels (Barcelona), Spain}
\author{S. Stringari}
\affiliation{BEC-INFM and Dipartimento di Fisica, Universita' di Trento, 
1-38050 Povo, Italy}

\begin{abstract}
We use Bogoliubov theory to calculate the beyond mean field correction 
to the equation of state of a weakly interacting Bose gas in the presence of a tight 2D 
optical lattice. We show that the lattice induces a characteristic 3D to 1D
crossover in the behaviour of quantum fluctuations.
Using the hydrodynamic theory of superfluids, we calculate the corresponding 
shift of the collective frequencies of a harmonically trapped gas.
We find that this correction can be of the order of a few percent
and hence easily measurable in current experiments. The behavior of the quantum depletion
of the condensate is also discussed.

\end{abstract}
\maketitle

Recently, the group of Ketterle at MIT reported \cite{kett2006} the first measurements of the 
quantum depletion in a condensate of $^{23}$Na atoms in tight optical lattices. 
The condensate fraction corresponds to the population
of the interference peaks observed in the time-of-flight images, whereas the 
remaining diffusive background is interpreted as the quantum depletion.

From the theoretical point of view, the quantum depletion of a weakly interacting Bose gas 
can be calculated from Bogoliubov theory. Another, closely related, quantity 
is the beyond mean field correction to the equation of state. In free space (no lattice)
this is known as the Lee-Huang-Yang (LHY) correction \cite{LHY}. In Ref.\cite{prl98}, 
Pitaevskii and Stringari
suggested to observe the LHY correction through the measurements of the collective frequencies
which are sensitive to changes in the equation of state. However, for tipical values of the atom
density and scattering length, this correction remains very small and difficult to measure. 
First experimental signatures of these corrections have been recently observed in a superfluid
 Fermi gas on the BEC side of the Feshbach resonance \cite{grimm}. 
The inclusion of the lattice is expected to enhance the role of correlations
and therefore effects beyond mean-field might become visible in such configurations.
Previous experimental \cite{esslinger} and theoretical \cite{chiara,paolo} attempts
to investigate beyond mean field effects in the collective frequencies of trapped Bose gases
have focused on quasi-1D systems, where the gas undergoes a crossover from a weakly interacting Bose 
Einstein condensate to a Tonks gas by decreasing the atomic density in the tube. 

In this Letter, we investigate a Bose-Einstein condensates in a tight 2D periodic potential
forming a 2D array of weakly coupled tubes. For a fixed atom density, the gas is in an anisotropic 3D 
regime at small values of the laser intensity and it undergoes a dimensional crossover to a quasi-1D
regime when the lattice depth is increased. 
We calculate the beyond-mean field correction to the ground state energy due to quantum fluctuations 
along the crossover.
 We then include the harmonic trapping and apply the hydrodynamic theory 
of superfluids to calculate the corresponding frequency shift of the 
lowest compressional mode along the axial direction, where atoms are free to oscillate subject to
the harmonic potential and two-body interactions.  
We show that for values of physical parameters available in current experiments, this shift 
becomes large enough to be detected. Our results in the asymptotic 1D regime are consistent with
those obtained in Ref.\cite{paolo}.

Let us first consider a Bose-Einstein condensate in the presence of a 
2D optical lattice
\be
V_{opt}(x,y)=s E_R [\sin^2(q_B x)+\sin^2(q_B y)],
\ee
where $s$ is the laser intensity, $E_R=\hbar^2 q_B^2/2m$ is the recoil energy, with $q_B$ the Bragg
quasi-momentum and $m$ the atomic mass. The lattice period is fixed by $d=\pi/q_B$.
Atoms are unconfined in the axial direction. The role of the additional harmonic trapping will be 
discussed later. 

The interparticle interaction is described by a s-wave contact potential with coupling constant 
$g=4\pi \hbar^2 a/m$, $a$ being the 3D scattering length.
We will discuss the situation where the laser intensity is sufficiently large $(s\gtrsim 5)$
and the chemical potential $\mu$ is small compared to the interband gap. We
thus restrict ourselves to the lowest Bloch band, where the physics is governed by the ratio
between the chemical potential and the bandwidth $8t$, where $t$ is the tunneling rate between 
neighboring wells. For $\mu \ll 8t$, the system retains an anisotropic 3D behaviour,
whereas for $\mu \approx 8t$, the system undergoes a dimensional crossover to a quasi-1D regime. 
Experimentally, this crossover can be realized by either increasing the laser intensity 
($t$ decreases) or increasing the atom density $n$ ($\mu $ increases).

In tight-binding approximation the states of the lowest Bloch band
can be written in terms of Wannier functions as $\phi_{k_x}(x)\phi_{k_y}(y)$, where
$\phi_k(u)= \sum_{\ell}e^{i\ell k u} w(u-\ell d)$ and $w(u) =\exp[-u^2/2\sigma^2]/\pi^{1/4}\sigma^{1/2}$
is a variational gaussian ansatz. By minimizing the free
energy functional with respect to 
$\sigma$, one finds  $d/\sigma\simeq \pi s^{1/4}\exp(-1/4\sqrt s)$.
The system Hamiltonian takes the form
\be\label{ham}
H^\prime=\sum_{\mathbf k} \epsilon_\mathbf k^0 a_\mathbf k^\dagger a_\mathbf k+
\frac{\tilde g}{2V}  
\sum_{\mathbf k,\mathbf q,{\mathbf k}^\prime} {a_{\mathbf k +\mathbf q}^\dagger}
{a_{\mathbf k^\prime-\mathbf q}^\dagger} a_{{\mathbf k}^\prime} a_\mathbf k, 
\ee
where $\epsilon_\mathbf k^0=k_z^2/2m+2t[2-\cos(k_x d)-\cos(k_y d)] $
is the energy dispersion of the non interacting model,
$V$ is the volume, and $\tilde g= C^2 g$ is an effective coupling constant, 
with $C=d \int_{-d/2}^{d/2} w^4(u) du\simeq d/\sqrt{2 \pi}\sigma$. 

In the following we assume that the number of atoms per tube is sufficiently large.
Under this assumption,  we can neglect the Mott 
insulator phase transition which would occur only for extremely large values of the 
laser intensity.  
Applying Bogoliubov theory to the Hamiltonian (\ref{ham}), 
the energy spectrum $E_\mathbf k $ of the elementary excitations is given by 
$E_\mathbf k^2=\epsilon_\mathbf k^0(\epsilon_\mathbf k^0+2 \tilde g n)$ \cite{javanainen}. 
The ground state energy of the system can be obtained from the relation
$E=n^2\tilde g V/2-\sum_{\mathbf k \ne 0} 
(\epsilon_\mathbf k^0+\tilde g n-E_\mathbf k)/2$, where the second term corresponds
to the beyond mean field correction due to quantum fluctuations. This
 correction can be shown to be always negative (in free space this formula would contain 
an ultraviolet  divergence to be cured by a proper renormalization of the coupling constant).

Replacing the sum with integrals and performing the integration 
over the axial momentum
$k_z$, we find
\be\label{key-en}
\frac{E}{V}=\frac{1}{2}\tilde g n^2-\frac{1}{4\pi} \frac{\tilde g n}{d^2}\sqrt{2m \tilde g n} f
\left(\frac{2t}{\tilde g n}\right),
\ee
where the function $f(x)$ is defined as
\be\label{f2}
f(x)=\frac{\pi}{2\sqrt{x}}\int_{-\pi}^{\pi}\frac{d^2\mathbf k}{(2\pi)^2} 
\frac{_2F_1\left[1/2,3/2,3,-\frac{2}{x \gamma(\mathbf k) }\right]}{\sqrt{\gamma(\mathbf k)}},
\ee
with $\gamma(\mathbf k)=2-\cos k_x-\cos k_y$. Here $_2F_1[a,b,c,d]$ is the hypergeometric function and  
the integration over the transverse quasi-momenta is restricted to the first Brillouin zone 
$|k_x|,|k_y|\leq \pi$. 
\begin{figure}[tb]
\begin{center}
\includegraphics[width=6.5cm,angle=270]{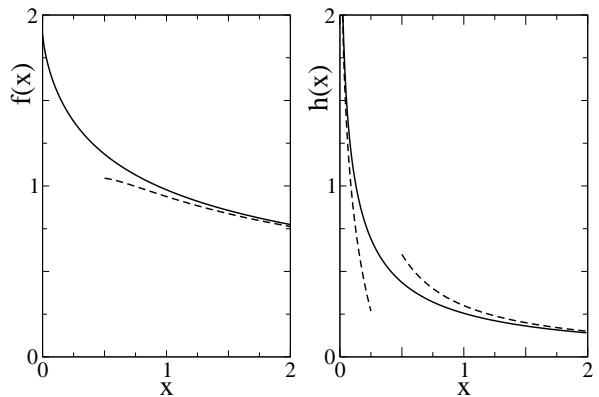}
\caption{Left panel: scaling function $f(x)$ [see Eq.(\ref{f2})](solid line) and its 
asymptotic behavior (dashed line). Right panel: scaling function $h(x)$ [see Eq.(\ref{nd})]
(solid line) and its asymptotic behavior (dashed lines).}
\label{homo}
\end{center}
\end{figure}
Equation (\ref{f2}) has been integrated numerically and the result is shown 
in Fig.\ref{homo} (left panel). For vanishing $x$, $f(x)$ saturates to 
the value $4 \sqrt{2}/3\simeq 1.89$. In this limit, corresponding to $8t \ll \mu$, we can neglect the 
Bloch dispersion and Eq.(\ref{key-en}) yields the ground state energy of a 1D Bose gas
\be\label{en1D}
\frac{E}{L}=\frac{1}{2} g_{1D} n_{1D}^2-\frac{2}{3\pi}\sqrt{m}(n_{1D} g_{1D})^{3/2},
\ee
with linear density $n_{1D}=n d^2$ and coupling constant $g_{1D}=\tilde g/d^2$. Here 
$L$ is the length of the tube.
Result (\ref{en1D}) is in agreement with the exact Lieb-Lininger solution \cite{lieb} 
of the 1D model expanded in 
the weak coupling regime $mg_{1D}/\hbar^2 n_{1D}\ll 1$.

In the opposite 3D regime $x\gg 1$, the function (\ref{f2}) approaches the asimptotic law 
$f(x)\simeq 1.43 /\sqrt{x}-16\sqrt{2}/15 \pi x$. This is shown in Fig.\ref{homo}
(left panel) with the dashed line. Introducing the notation $\tilde a=C^2 a$, 
Eq.(\ref{key-en}) takes the asymptotic form
\be\label{en3D}
\frac{E}{V}=\frac{2\pi}{m}n^2 \tilde a \left(1+\frac{\tilde a}{\tilde a_{cr}}+
\frac{128}{15}\left(\frac{n\tilde a^3}{\pi}\right)^{1/2} \frac{m^*}{m}\right),\;\;\;
\ee 
where $m^*=\hbar^2/2t d^2$ is the effective mass associated to the band and
$\tilde a_{cr}=-0.24 d \sqrt{m/m^*}$.
The last term in the rhs of Eq.(\ref{en3D}) corresponds to
the generalized LHY correction in the presence of 
the optical lattice. We see that with respect to the free case, the LHY correction
is magnified  by the renormalization of both the coupling constant 
$(\tilde a>a)$ and the effective mass $(m^*>m)$. It is worth noticing that
Eq.(\ref{en3D}) is valid only for $\mu\ll 8t$, a condition which, for large $s$,  requires
ultra-low atomic densities.

The term $\tilde a/\tilde a_{cr}<0$ in Eq.(\ref{en3D}) is more 
subtle and amounts to a further 
renormalization of the scattering length due to the optical lattice.
In particular, the correct value for the scattering length for low-energy ($E \ll 8t)$
2-body collisions is given by  $1/a_{eff}=1/\tilde a-1/\tilde a_{cr}$ \cite{pra}.
In Eq.(\ref{en3D}) $a_{eff}$ is replaced by the linear expansion
$a_{eff} \simeq \tilde a(1+\tilde a/\tilde a_{cr})$.

The equation of state $\mu=\partial E/\partial N$ can be obtained by differentiating 
Eq.(\ref{key-en}). We find it convenient to write it as $\mu=\tilde g n[1+k(n)]$, where 
the term proportional to $k(n)$ accounts for the effects of quantum fluctuations.
In the quasi-1D limit $k(n)=-1/(\pi d^2)\sqrt{m\tilde g/n}$ whereas in 
the opposite 3D regime we find $k(n)=\tilde a/\tilde a_{cr}+\beta (m^*/m) \sqrt{\tilde a^3 n}$, 
with $\beta=32/3\sqrt{\pi}$.
  
Using the same formalism it is possible to calculate the quantum depletion
given by $N-N_0=\sum_{\mathbf k \ne 0} (\epsilon_\mathbf k^0+\tilde g n-
E_\mathbf k)/2 E_\mathbf k$. 
Taking the continuum limit, we find
\be\label{nd}
\frac{N-N_0}{N}=\frac{1}{4\pi}\frac{1}{d^2}\sqrt{\frac{2m \tilde g}{ n}}
h\left(\frac{2t}{\tilde g n}\right),
\ee
with
\be\label{def_h}
h(x)=\int_{-\pi}^{\pi}\frac{d^2\mathbf k}{(2\pi)^2}
\int_{x \gamma(\mathbf k)}^{\infty} \frac{dt}{\sqrt{t-x\gamma(\mathbf k)  }}\left[\frac{t+1}{\sqrt{t^2+2t}}-1\right].
\ee
Differently from the beyond mean field correction to the ground state energy (\ref{en1D}), 
the quantum depletion diverges for vanishing tunneling  
as $h(x)\simeq -\ln(2.7 x)/\sqrt{2}$. This signals that in the 
absence of tunneling 
there is no real Bose-Einstein condensation in agreement with the general theorems in 1D \cite{book}. 
In the opposite 3D regime $x\gg 1$ the function $h(x)$ decays as $4/(3\pi \sqrt{2}x)$ and from
Eq.(\ref{nd}) we find $(N-N_0)/N=8m^*/(3m) (\tilde a^3 n/\pi)^{1/2}$, which generalizes the 
standard 3D result in free space \cite{book}. 

In the second part of the Letter
we apply the hydrodynamic theory of superfluids to investigate the effects
of quantum correlations on the collective frequencies of a trapped gas.
Expanding the atom density as $n(\mathbf r,t)=n(\mathbf r)+\delta n(\mathbf r,t)$, the 
hydrodynamic equations in the presence of the lattice can be written in the useful form 
\cite{meret} 
\be
\label{hhd1}
m{\partial^2 \delta  n \over \partial t^2} - 
\tilde {\mbox{\boldmath$\nabla$}}\left[n \tilde{\mbox{\boldmath$\nabla$}}
\left(\frac{\partial \mu}{\partial n}\delta n \right)\right] = 0 \, ,
\ee
where we have introduced the notation 
$\tilde {\mbox{\boldmath$\nabla$}}\equiv (\mbox{\boldmath$\nabla$}_\perp \sqrt{m/m^*},\nabla_z)$.
For trapped configurations, the ground state density $n(\mathbf r)$ entering Eq.(\ref{hhd1})
has to be calculated by imposing the local equilibrium condition
$\mu_0=\mu(n(\mathbf r))+V_{ext}(\mathbf r)$, where $\mu_0$ is the ground state value of the 
chemical potential, fixed by the proper normalization of $n(\mathbf r)$, and 
$V_{ext}(\mathbf r)=(\omega_\perp^2 \mathbf r_\perp^2+\omega_z z^2)/2 m $ is the 
external trapping potential, here assumed of axial symmetry.

By substituting $\mu=\tilde g n[1+k(n)]$ into Eq.(\ref{hhd1}) and retaining only 
terms linear in $k(n)$, we obtain
\begin{equation}
m \omega^2\delta n + 
\tilde{\mbox{\boldmath$\nabla$}} \left(gn_{TF} \tilde{\mbox{\boldmath$\nabla$}}
\delta n\right) = - \tilde{\mbox{\boldmath$\nabla$}}^2\left(\tilde g  
n_{TF}^{2}\frac{\partial k }{\partial n_{TF}}  \delta n\right),
\label{HD1}
\end{equation}
where $n_{TF}(\mathbf r)=[\mu_0-V_{ext}(\mathbf r)]/\tilde g$ is the 
Thomas-Fermi density profile.
Equation (\ref{HD1}) can be solved by treating its right hand side as a 
small perturbation, following the procedure of Ref.\cite{prl98}.
To this purpose one first solves the associated
zero-th order hydrodynamic equation setting the rhs of Eq.(\ref{HD1}) equal to zero \cite{sandro96}.
The corresponding solution can then be used to calculate the frequency shift
induced by the perturbation. One finds
\begin{equation}
\frac{\delta \omega}{\omega} = -\frac{\tilde g}{2 m\omega^2}
\frac{\int d^3{\bf r} \delta n (\tilde{\mbox{\boldmath$\nabla$}}^2 \delta n^*) 
n_{TF}^{2}\partial k/\partial n_{TF}}
{\int d^3{\bf r}  \delta n^*\delta n},
\label{deltaomega}
\end{equation}
where the integrals extend to the region where the Thomas-Fermi density is positive.
We see from Eq.(\ref{deltaomega}) that the shift in the frequency is not proportional
to $k(n)$ but rather to its derivative $\partial k/\partial n$, so that the
{\sl density-independent} term $\tilde a/\tilde a_{cr}$ does not 
contribute to the frequency shift (\ref{deltaomega}). Furthermore,
the shift can be positive even if $k(n)$ is negative, as it happens in the quasi 1D regime. 

In order to observe effects beyond mean field, we focus on the lowest compressional 
mode along the axial direction. For simplicity we assume an {\sl effective} disc-shaped trap 
$\omega_\perp/\sqrt{m^*} \ll \omega_z$. In this case the zero-th order dispersion is given
by $\omega=\sqrt{3}\omega_z$ with density oscillations of the form 
$\delta n(\mathbf r) \sim z^2-Z_{TF}^2/3$, where  $Z_{TF}^2=2 \mu/m\omega_z^2$ is the square of the 
Thomas-Fermi radius along the axial direction.   
Calculating $k(n)$ from Eq.(\ref{key-en}) and inserting it 
into Eq.(\ref{deltaomega}), after integration over coordinates we find  
\be\label{finalomega}
\frac{\delta \omega}{\omega}=\frac{21}{512} \sqrt{\frac{m\tilde g}{n(0)}}\frac{1}{d^2}Y\left(
\frac{2t}{\tilde g n(0)}\right),
\ee
where $n(0)=n_{TF}(0)$ is the density evaluated in the center of the trap and 
\be\label{ygrande}
Y(x)=\frac{16 \sqrt{2}}{\pi} \sqrt{x} \int_0^1 dr r^2(r^2-1) G\left(\frac{x}{1-r^2}\right).
\ee
Here $x=2t/\tilde g n(0)$ is the parameter controlling the dimensional crossover for {\sl trapped}
gases and $G(y)=y^{3/2} f^{\prime \prime}(y)-3 f(y)/4\sqrt{y}$. 
The function (\ref{ygrande}) satisfies $Y(0)=1$ and decreases monotonically as $x$ increases, reaching the 
asymptotic law $Y(x)=5/6 \pi x$ for large values of $x$.

Equations (\ref{finalomega})-(\ref{ygrande}) are the key results of this Letter.
In the quasi-1D regime, corresponding to large values of the laser intensity,
$\delta \omega/\omega=(21/512 )\sqrt{m \tilde g/n(0)d^4} $, showing that 
the shift increases by decreasing the central density.
In the opposite 3D regime of small values of $s$, the frequency shift is instead given by 
$\delta \omega/\omega=(35 \sqrt{\pi} m^*/ 128 m) \sqrt{\tilde a^3 n(0)},
$
and hence increases by increasing the central density. This latter equation 
provides the lattice generalization of the frequency shift  
due to the LHY correction investigated in Ref.\cite{prl98}.

\begin{figure}[tb]
\begin{center}
\includegraphics[width=6.5cm,angle=270]{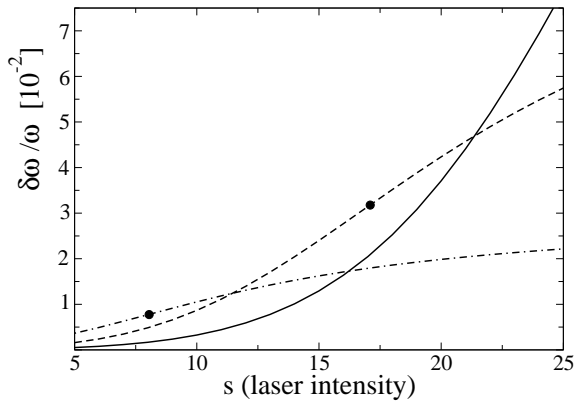}
\caption{Calculated frequency shift of the axial mode for a bosonic cloud of $^{23}$Na atoms 
in a disc trap as a function of the laser intensity for different values of the central density 
$n(0)=10^{12}$cm$^{-3}$ (solid line),
$10^{13}$cm$^{-3}$ (dashed line) and $10^{14}$cm$^{-3}$ (dashed-dotted line).
The circles correspond to $x=0.25$, when the chemical potential  
in the center of the trap is equal to the bandwith [for $n(0)=10^{12}$cm$^{-3}$ the 
1D regime is achieved for larger values of $s$].
Parameters used: $a=2.75$ nm, $d=297.3$ nm.}
\label{kett}
\end{center}
\end{figure}

In Fig.\ref{kett} we show the calculated frequency shift as a function of the laser intensity 
for atomic samples of $^{23}$Na (3D scattering length $a=2.75$ nm) in a 2D optical 
lattice with period $d=297.3$ nm (generated by a dye laser) for different values of the
central density. 
We see that the frequency shift can be of the order of a few  percent for  
values of $s$  easily accessible in experiments. 

\begin{figure}[tb]
\begin{center}
\includegraphics[width=6.5cm,angle=270]{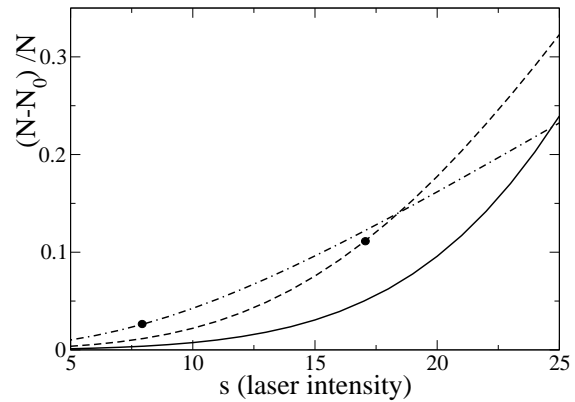}
\caption{Calculated quantum depletion for a bosonic cloud of $^{23}$Na atoms 
in a disc trap as a function of the laser intensity for different values of the central density 
$n(0)=10^{12}$cm$^{-3}$ (solid line),
$10^{13}$cm$^{-3}$ (dashed line) and $10^{14}$cm$^{-3}$ (dashed-dotted line).
Parameters as in Fig.\ref{kett}.}
\label{kett-qd}
\end{center}
\end{figure}

It is interesting to compare the frequency shift with the quantum depletion 
in the presence of the trap. Starting from Eqs (\ref{nd}) and (\ref{def_h}), and inserting 
the external potential via a local density approximation, we find
\be
\label{quantum-dep}
\frac{N-N_0}{N}=\frac{15 \sqrt{2}}{8\pi} \sqrt{\frac{m\tilde g}{n(0)}}\frac{1}{d^2}
Q\left(\frac{2t}{\tilde g n(0)}\right),
\ee
where the function $Q(x)$ is defined as
\be
Q(x)=\int_0^1 dr r^2 (1-r^2)^{1/2} h\left(\frac{x}{1-r^2}\right).
\ee
For vanishing $x$, $Q(x)$ diverges logarithmically 
as $Q(x)=-0.26-\pi \ln x/16 \sqrt{2} $ whereas in the 3D regime $x \gg 1$ we find 
$Q(x) =1/24 \sqrt{2}x$, yielding the known result $(N-N_0)/N=(5\sqrt{\pi}m^*/8m)\sqrt{\tilde a^3 n(0)}$
\cite{book}. 
In Fig.\ref{kett-qd} we plot the quantum depletion evaluated from Eq.(\ref{quantum-dep}) 
as a function of the laser intensity for the same parameters used in Fig.\ref{kett}.
The comparison with Fig.\ref{kett} shows that the effects of the lattice on the quantum depletion 
are larger than on the frequency shifts. However, one should remind that the measurement
of the collective frequencies can be obtained wich much higher precision. 

Let us finally discuss the conditions of applicability of our results. First, the effects of the 
trap are taken into account via local density approximation. This requires
that the trapping frequencies $\omega_\perp ,\omega_z$ are small compared to the chemical potential
$\mu$ and the bandwidth $8t$. In particular the condition $\omega_z \ll 8t$ ensures that
the gas oscillating along the tubes retains the 3D coherence.
Second, the mean-field value $\omega=\sqrt{3}\omega_z$ for the frequency of the collective
oscillation is only valid for a strongly anisotropic trap $\omega_\perp/\sqrt{m^*} \ll \omega_z$.
For a finite value of the anisotropy, one should start from the more general formula
obtained in Ref.\cite{prl98} valid for an arbitrary anisotropy in the absence of the lattice. 
We see from Eq.(\ref{hhd1}) that in the hydrodynamic theory 
the lattice enters through the renormalization of the effective mass $m\rightarrow m^*$
along the confined directions. From Ref.\cite{prl98} we then obtain 
$(\delta \omega/\omega)_\textrm{ani}=(m/ 9 m^*) \omega_\perp^2/\omega_z^2$,
showing that 
the correction {\sl decreases} by increasing the laser intensity, being proportional to $m/m^*$.

In conclusions, our results show that the measurement of the
collective frequencies in the presence of 2D optical lattices can provide an efficient tool to 
investigate beyond mean field effects along the 
dimensional crossover. 

We acknowledge interesting discussions with L. Pitaevskii and M. Wouters.
This work was supported by the Ministero dell'Istruzione, dell'Universita' e della Ricerca (M.I.U.R.).

\end{document}